\documentclass{elsart}
\usepackage{graphicx}

\begin{document}

\begin{frontmatter}

\title{Pairing transition of nuclei at finite temperature}

\author[Higashi]{K. Kaneko} and \author[Sawara]{M. Hasegawa}
\address[Higashi]{Department of Physics, Kyushu Sangyo University,
 Fukuoka 813-8503, Japan}
\address[Sawara]{Laboratory of Physics, Fukuoka Dental College,
 Fukuoka 814-0193, Japan}

\begin{abstract}
Pairing transition at finite temperature was investigated by the shell model 
and BCS calculations. 
The definitive signature of pairing transition is identified by a ``transition temperature" 
$T_t$ estimated from a ``thermal" odd-even mass 
difference, while there is no sharp phase transition because of the finiteness of nucleus. 
It is found that $T_t$ is in good agreement with predictions of critical temperature 
$T_c$ in the BCS approximation, 
and the pairing correlations almost vanish at 
two points of the transition temperature $T\approx 2T_{t}$. 
The BCS calculations show that the critical 
temperature $T_c$ increases with increasing deformation. 

\vspace*{3mm}
\leftline{PACS:  21.60.Cs, 21.10.Ma, 05.30.-d}
\end{abstract}

\end{frontmatter}

\section{Introduction}

Pairing correlations are one of the fundamental properties of nuclei. 
The odd-even mass difference observed in nuclear masses, and the energy gap of 
even-even nuclei are well known as signatures of pairing effects \cite{Bohr}.  
The Bardeen-Cooper-Schriffer (BCS) theory \cite{BCS} of superconductivity has been applied to 
such nuclear problems at zero temperature \cite{Belyaev}. 
The thermodynamical properties of nuclear pairing 
were investigated using the BCS theory in the study of hot nuclei \cite{Sano,Goodman}. 
Breaking of the Cooper pairs is expected to occur above a certain critical temperature. 
Infinite Fermi systems show 
a sharp phase transition from a super-fluid phase to a normal-fluid one at a critical 
temperature $T_{c}$, where the 
pairing gap vanishes and the heat capacity exhibits a singularity. 
For metal superconductors, the critical temperature is $T_{c}\approx 0.57\Delta$ MeV 
\cite{BCS}, 
where the pairing gap $\Delta$ is calculated using the BCS equation. 
In this case, we can image a well deformed nucleus for which level density is 
almost uniform and average level spacing is very small compared with the pairing gap $\Delta$ 
($\approx$ 1 MeV for the rare earth nuclei) \cite{Ring}. 
On the other hand, the pairing correlations vanish at 
$T_{c}\approx 0.50\Delta$ MeV \cite{Goodman1} in a simple pairing model with half-filled 
degenerate shells. 

However in the case of the finite Fermi system like a nucleus, since the nuclear radius is much 
smaller than the coherence length of the Cooper pair, statistical fluctuations beyond the mean field 
become larger. The fluctuations smooth out the sharp phase transition, and 
then 
the pairing correlations do not 
vanish but decrease with increasing temperature. 
There are many approaches for treating the fluctuations beyond the mean field. 
A signature of the pairing transition at finite temperature might be 
a peak of the heat 
capacity as a function of temperature \cite{Rossignoli}. 
In fact, it has recently been reported \cite{Schiller,Melby} that the canonical heat capacities 
extracted from observed level densities in $^{162}$Dy, $^{166}$Er and $^{172}$Yb
form an S shape 
around $T\approx 0.5$ MeV, which is interpreted as the breaking of nucleon 
Cooper pairs and the pairing transition 
because the critical temperature corresponds to 
$T_{c}\approx 0.57\Delta\approx 0.5$ MeV in the BCS theory. 
The odd-even effect that shows that the heat capacity of 
an odd-mass nucleus is smaller than those of the adjacent even-mass nuclei can be found 
around the critical temperature $T_{c}$. 

The spherical shell model approaches could be more appropriate for describing various 
aspects of nuclear structure. The large fluctuations can be taken into account 
beyond the mean field in the shell model calculations. For the description of 
nuclear properties at higher temperatures, one needs a large model space. 
Recently the shell model Monte Carlo (SMMC) calculations have been performed
with the pure pairing force \cite{Rombouts} in a large model space,
and with the pairing and multipole-multipole forces \cite{Liu} in
the $fp + g_{9/2}$ shell model space for the even- and odd-mass Fe isotopes.
 It was shown that the pairing correlations would be important only at low 
temperatures and at low excitation energies \cite{Rombouts},
and that the suppression of pairing correlations due to finite temperature
appears as the S shape of the heat capacity around the temperature
$T\sim 0.8$ MeV \cite{Liu}. 
The model space was quite recently extended to examine the partition functions
and the level densities up to the higher temperature $T=4.0$ MeV
using an independent-particle 
approximation \cite{Alhassid} combined with the SMMC method. 

This paper is organized as follows. 
In Section 2, we carry out the shell model calculations 
in the large model space $(sd + fp+ s_{1/2}d_{5/2})$ using 
a spherical Woods-Saxon potential \cite{Cwoik}, where 
we adopt the independent-particle approximation \cite{Alhassid} 
and combine it with the shell model calculations in the $sd$-shell, 
though there are no contributions from many-body correlations
in the $fp$ shell or from the coupling between 
the $fp$ and $sd$ shells. 
The pairing correlations within the heat capacity 
or ``thermal" odd-even mass difference as a function of finite temperature are estimated, 
and a ``transition temperature" $T_t$ is introduced to explain the pairing transition. 
Then the definitive signatures of the pairing transition are identified. 
The critical temperature $T_{c}$ is also evaluated using the BCS calculations with 
the axially deformed Woods-Saxon potential and a pairing residual interaction in Section 3. 
We discuss the relations between the transition temperature $T_{t}$ and the critical 
temperature $T_{c}$. 
The systematic behavior of the critical temperature $T_{c}$ is examined over a wide range 
of nuclei. We discuss the effects of deformation on the 
critical temperature. 
Concluding remarks are given in Section 4. 

\section{Pairing transition in shell model calculations}

We start from the canonical partition function defined by 
\begin{eqnarray}
Z(T) = {\rm Tr}({\rm e}^{-H/T}) = \sum_{i=0}^{\infty}{\rm e}^{-E_{i}/T}, 
\label{eq:1}
\end{eqnarray}
where $E_{i}$ is the energy of the $i$th eigenstate for the Hamiltonian $H$ 
of a system. 
The large matrix of the Hamiltonian $H$ is diagonalized to obtain all the eigenvalues $E_{i}$, 
and the partition function in the canonical ensemble is calculated from Eq. (\ref{eq:1}). 
Then, any thermodynamical quantities $O(T)$ can be evaluated from 
\begin{eqnarray}
O(T) = \langle O \rangle = {\rm Tr}(O{\rm e}^{-H/T})/Z(T), 
\label{eq:2}
\end{eqnarray}
where $\langle O \rangle$ stands for the average value of operator $O$ over the range of 
eigenstates. For instance, the thermal energy is expressed as
\begin{eqnarray}
E(Z,N,T) = \langle H \rangle = \sum_{i=0}^{\infty}E_{i}{\rm e}^{-E_{i}/T}/Z(T). 
\label{eq:3}
\end{eqnarray}
The heat capacity is then given by 
\begin{eqnarray}
C(Z,N,T) = \frac{\partial E(Z,N,T)}{\partial T}.
\label{eq:4}
\end{eqnarray}

We now introduce the thermal odd-even mass difference for neutrons defined by 
the following three-point indicator: 
\begin{eqnarray}
\Delta_{n}^{(3)}(Z,N,T) & = & \frac{(-1)^{N}}{2}[ E(Z,N+1,T) \nonumber \\
   & {} & - 2E(Z,N,T)+E(Z,N-1,T)]. 
\label{eq:5}
\end{eqnarray}
The odd-even mass difference at zero temperature is known theoretically and experimentally 
as an important quantity in evaluation of the pairing correlations in a nucleus. The thermal 
odd-even mass difference is also an indicator of the pairing correlations at 
finite temperature, and can be obtained from the experimental energies and the level 
density as well as the heat capacity. 
\begin{figure}[t]
\begin{center}
\includegraphics[width=8cm,height=10cm]{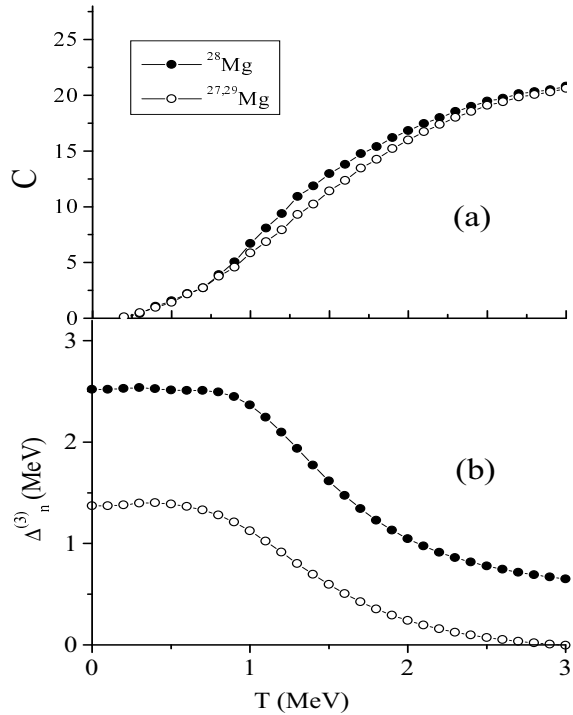}
  \caption{Heat capacity (a) and thermal odd-even mass difference (b) 
  as a function of temperature. The solid circles denote the values for even-even nucleus 
  $^{28}$Mg. The open circles denote the average values of the neighboring odd nuclei 
  $^{27}$Mg and $^{29}$Mg. 
  }
  \label{fig1}
\end{center}
\end{figure}

\begin{figure}[t]
\begin{center}
\includegraphics[width=8cm,height=8cm]{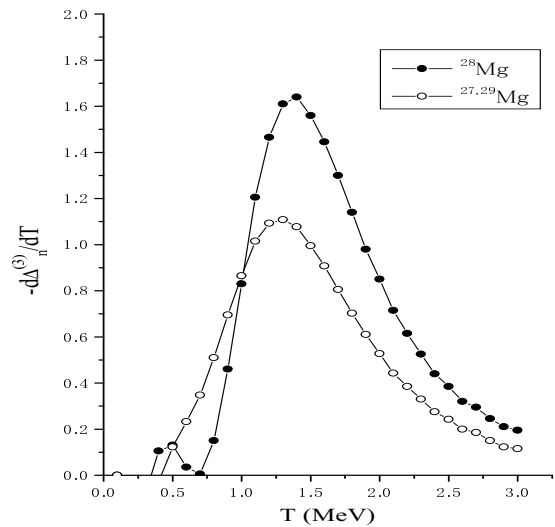}
  \caption{Derivative of the thermal odd-even mass difference.}
  \label{fig2}
\end{center}
\end{figure}

We now calculate the heat capacity (\ref{eq:4}) and the thermal odd-even mass difference 
(\ref{eq:5}) for 
$^{20-26}$Ne and $^{24-32}$Mg in the large model space $(sd + fp+ s_{1/2}d_{5/2})$ 
using an independent-particle approximation \cite{Alhassid}. 
The Woods-Saxon potential plus a spin-orbit interaction is diagonalized in a basis 
of harmonic-oscillator (H.O.) eigenfunctions, and then 
the single-particle energies are obtained \cite{Cwoik}. 
The Woods-Saxon parameters are chosen so as to reproduce the single-particle energies 
estimated from $^{17}$O. 
A number of unbound states ($E>0$) are obtained due to the expansion in a finite 
number of H.O. eigenfunctions. 
The resonances with narrow widths are important for neutron-rich nuclei 
such as $^{66}$Cr. 
It was shown \cite{Alhassid}, however, that while the narrow resonances
make an important contribution to the partition function, 
their width can be ignored. In this paper, 
we ignore the widths of resonances and the continuum states. 
We carried out the shell model calculations
in the $sd$-shell model space using the USD interaction \cite{USD},
 and calculated the correlated thermal energy $E_{v,tr}$
 using Eq. (\ref{eq:3}). 
An enhancement of the heat capacity was found in even Mg nuclei around $T\sim$ 1.5 MeV. 
This enhancement is interpreted as a reduction in the pairing correlations. However, 
because the calculation was restricted to a finite space ($sd$ shell), the calculated 
heat capacity reached its maximum around temperatures of $\sim$ 1.5 MeV. 
Therefore, we extended 
the model space to $sd + fp + s_{1/2}d_{5/2}$ to display a much broader temperature range 
($\sim$ 3.5 MeV). 
We combine the small model space $sd$ of interaction effects with a large-space calculation 
of the independent-particle thermal energy $E_{sp}$ as follows \cite{Alhassid}:
\begin{eqnarray}
E & = & E_{v,tr} + E_{sp} - E_{sp,tr},
\label{eq:6}
\end{eqnarray}
where $E_{sp,tr}$ is the independent-particle thermal energy in the $sd$ space. 
Then the heat capacity and the thermal odd-even mass difference are 
obtained using Eqs. (\ref{eq:4}) and (\ref{eq:5}), respectively. 
As typical examples, the heat capacity and the thermal odd-even mass difference in 
$^{27-29}$Mg are plotted in Fig. \ref{fig1} (a) and (b), 
where the average values of $^{27}$Mg and $^{29}$Mg are 
shown for odd-mass nuclei. The heat capacity $C(^{28}{\rm Mg})$ is 
larger than the average value $C(^{27,29}{\rm Mg})=[C(^{27}{\rm Mg})+C(^{29}{\rm Mg})]/2$ 
in $T = 0.8 - 3.0 $ MeV, 
although it does not reach a peak. 
In other nuclei, we also found an odd-even effect where the average heat capacity of 
odd-mass nuclei is significantly lower than that of the adjacent even-mass nucleus. 
In Fig. \ref{fig1} (a), one can see that 
with the usual relation $C=2aT$ 
the parameter $a$ has roughly the empirical 
value of $A/8$ MeV$^{-1}$ \cite{Bohr1} for light nuclei, and is considerably larger than 
the Fermi-gas model value $A/15$ MeV$^{-1}$. 
Figure \ref{fig1} (b) shows the thermal odd-even mass difference defined by 
Eq. (\ref{eq:5}) for $^{28}$Mg and $^{27,29}$Mg. We find a gradual decrease of the thermal 
odd-even mass difference as a function of temperature, which is interpreted as a gradual breaking 
of nucleon Cooper pairs and the decline of pairing correlations. 
Figure \ref{fig1} (b) also displays that $\Delta_{n}^{(3)}(^{28}{\rm Mg})$ is larger 
than $\Delta_{n}^{(3)}(^{27,29}{\rm Mg})=[\Delta_{n}^{(3)}(^{27}{\rm Mg})+
\Delta_{n}^{(3)}(^{29}{\rm Mg})]/2$. At zero temperature, this is well known as 
the odd-even staggering of binding energies, which reflects the stronger binding of 
even-particle-number systems than the odd-particle-number neighbors \cite{Satula,Dobaczewski}. 
Their analyses demonstrated that the odd-even mass difference for 
odd-particle number is an excellent measure of pairing correlations, 
although it is still controversial \cite{Duguet}. 
The symmetric filter 
$\delta e=2\Delta_{n}^{(3)}(2m)-\Delta_{n}^{(3)}(2m-1)-\Delta_{n}^{(3)}(2m+1)$ 
extracts the effective single-particle 
spacing from the measured binding energies of deformed nuclei. We can see that the difference, 
$\delta e/2$, between $\Delta_{n}^{(3)}(^{28}{\rm Mg})$ and $\Delta_{n}^{(3)}(^{27,29}{\rm Mg})$ 
decreases gradually as temperature increases. This means that the effective single-particle 
spacing decreases as temperature increases. 
As one can see in Fig. \ref{fig1} (b), $\Delta_{n}^{(3)}(^{27,29}{\rm Mg})$ is almost 
zero at $T\approx 3.0$ MeV, and the pairing correlations vanish at two points on the transition 
temperature $T_{t}$. 
On the other hand, $\Delta_{n}^{(3)}(^{28}{\rm Mg})$ still remains at this 
temperature. 
We now identify an inflection point of the curve $\Delta_{n}^{(3)}(^{27,29}{\rm Mg})$ 
in Fig. \ref{fig1} (b) 
as a signature of pairing transition, and call it ``transition temperature" $T_{t}$. 
To see more precise position of the inflection point, we differentiate 
$\Delta_{n}^{(3)}(^{27,29}{\rm Mg})$ with respect to temperature $T$. 
As seen in Fig. \ref{fig2}, $-\partial\Delta_{n}^{(3)}(^{27,29}{\rm Mg})/\partial T$ 
has a peak at temperature $T_{c}\approx 1.3$ MeV corresponding to the transition 
temperature $T_{t}$, and 
the transition temperature $T_{t}$ for $^{27,29}{\rm Mg}$ is quite close to that for 
$^{28}{\rm Mg}$. 
It is very important to note 
that the difference between the two curves of the heat capacities 
in Fig. \ref{fig1} (a) is equal to $-\partial\Delta_{n}^{(3)}(^{28}{\rm Mg})/\partial T$ 
\begin{eqnarray}
-\frac{\partial\Delta_{n}^{(3)}(Z,N,T)}{\partial T} & = & (-1)^{N}\{ C(Z,N,T) - 
\frac{1}{2}[C(Z,N+1,T) \nonumber \\
    & {} & +C(Z,N-1,T)] \}. 
\label{eq:7}
\end{eqnarray}
Thus, the thermal odd-even mass difference is a good indicator for the pairing transition 
at the transition temperature estimated from a peak of $-\partial\Delta_{n}^{(3)}/\partial T$. 
It is important to note that the position of the peak does
 not change upon extension of the model 
space in the independent-particle approximation, which is 
 different from the result of Ref. \cite{Rombouts}. 

\begin{figure}[b]
\begin{center}
\includegraphics[width=8cm,height=8cm]{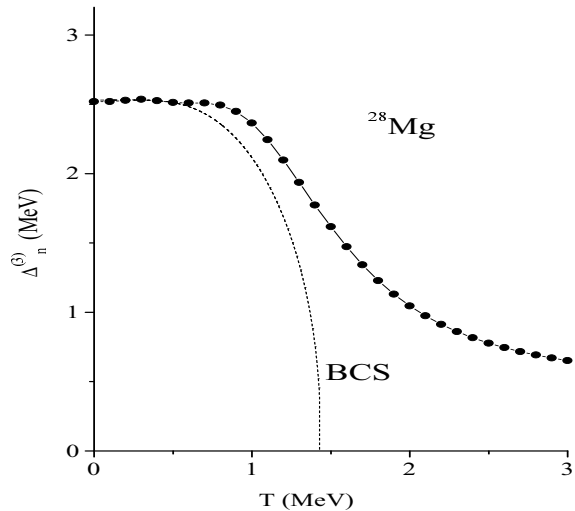}
  \caption{Thermal odd-even mass difference. 
   The solid circles represent the even-even nucleus 
   $^{28}$Mg, and the dotted line the pairing gap in the BCS 
   approximation.}
  \label{fig3}
\end{center}
\end{figure}
\begin{figure}[t]
\begin{center}
\includegraphics[width=8cm,height=8cm]{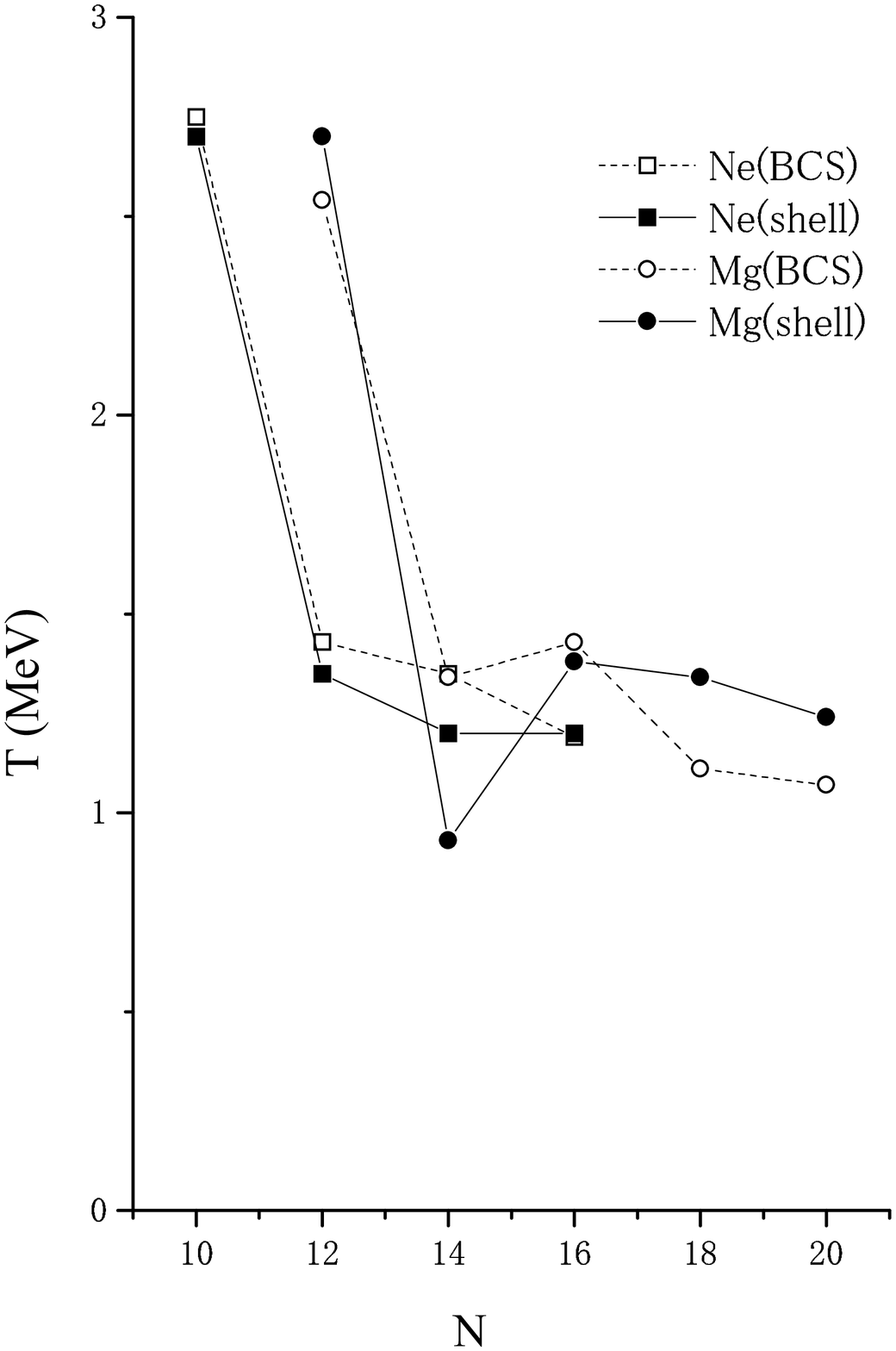}
  \caption{Transition temperature $T_{t}$ (solid symbols) and 
  critical temperature $T_{c}$ (open symbols) as a function of neutron number 
  in the Ne and Mg isotopes.}
  \label{fig4}
\end{center}
\end{figure}
\section{BCS calculations}

In infinite systems, the BCS critical temperature 
is known to be proportional to the pairing gap $T_{c} = 0.57\Delta$.
However, the relation is not axiomatic in a finite system like a nucleus. 
It seems that the peaks of the S-shaped heat capacities in 
$^{162}$Dy, $^{166}$Er, and $^{172}$Yb correspond to the critical temperature 
$T_{c}= 0.57\Delta$ in the BCS prediction for infinite systems. 
It is therefore interesting to examine the correspondence 
using the BCS calculations over a wide range of nuclei. 
In the BCS approximation, the pair gap $\Delta$ 
at finite temperature is obtained by using the following gap equation 
\begin{eqnarray}
1 & = & G\sum_{k>0}\frac{1-2f_{k}}{E_{k}},
\label{eq:6}
\end{eqnarray}
where $\varepsilon_{k}$ are the single-particle energies, 
$E_{k}=\sqrt{(\varepsilon_{k}-\mu)^{2}+\Delta^{2}}$ 
the quasiparticle energies, and $f_{k}=(1+{\rm e}^{E_{k}/T})^{-1}$ 
the Fermi-Dirac quasiparticle occupancies.
The chemical potential $\mu$ is determined by the number constraint
\begin{eqnarray}
N & = & \sum_{k>0}[1 - 
\frac{\varepsilon_{k}-\mu}{E_{k}}(1-2f_{k})],
\label{eq:7}
\end{eqnarray}
where the pairing force strength $G$ is chosen so as to reproduce the experimental 
odd-even mass difference at zero temperature. 
In this calculation, we use the single-particle energies extracted from 
an axially deformed Woods-Saxon potential with spin-orbit interaction \cite{Cwoik}. 
We chose the parameters so as to accommodate experimental single-particle 
energies extracted from energy levels of odd nucleus with double-closed core 
plus one neutron, i.e., $^{41}$Ca, $^{101}$Sn, and 
$^{209}$Pb. 
The deformation takes into account the effects of a quadrupole-quadrupole 
interaction in the mean-field approximation. 
The deformation parameter in even-even nuclei 
can be estimated from $B(E2)=[(3/4\pi)Zer_{0}^{2}A^{2/3}\beta]^{2}$ using experimental 
$B(E2)$ values. If they are not available, we adopt the empirical formula 
$B(E2)\approx 3.27E_{2_{1}^{+}}^{-1.0}Z^{2}A^{-0.69}$ for the $B(E2)$ values 
where $E_{2_{1}^{+}}$ is the energy of first excited $2^{+}$ state in even-even nucleus. 
The number of two-fold degenerate active orbitals is chosen as 20 corresponding to 
N=2,3 shells above $^{40}$Ca 
core in $sd$-nuclei, and 20 corresponding to N=3,4 shells above $^{100}$Sn core in 
$fp$-nuclei for each proton and neutron. 
For heavy nuclei, the numbers of active proton 
and neutron orbitals are taken as 30 corresponding to N=4,5 and N=5,6 shells above 
$^{208}$Pb core, respectively. 
The BCS pairing gap for $^{28}$Mg 
is shown in Fig. \ref{fig3}. The value of $\Delta_{n}^{(3)}$ 
decreases with increasing temperature, and vanishes at 
$T_{c}\approx 1.4$ MeV 
for $^{28}$Mg. We can see that 
the critical temperature for $^{28}$Mg is very close to the peak of the heat 
capacity. 

For the Ne and Mg isotopes examined, we examine the temperature about the peak of 
$-\partial\Delta_{n}^{(3)}/\partial T$ in the shell model calculation and the critical 
temperature $T_{c}$ in the BCS. 
Figure \ref{fig4} suggests that 
the critical temperature $T_{c}$ can be identified with the transition 
temperature $T_{t}$ even though the pairing correlations do not vanish. 
It is very important to note that the $N=Z$ nuclei $^{20}$Ne, $^{24}$Mg show a high 
transition temperature $T_{t}\approx$ 2.7 MeV, while the other $N>Z$ nuclei  
have $T_{t}\approx$ 1.3 MeV. In our previous paper \cite{kaneko}, 
we suggested that proton-neutron ({\it p-n}) correlations 
give rise to large odd-even mass difference in $N=Z$ nuclei. 
The cooperation of the {\it p-n} pairing correlations 
with the like-nucleon correlations would increase the transition 
temperature as well. 
\begin{figure}[t]
\begin{center}
\includegraphics[width=8cm,height=10cm]{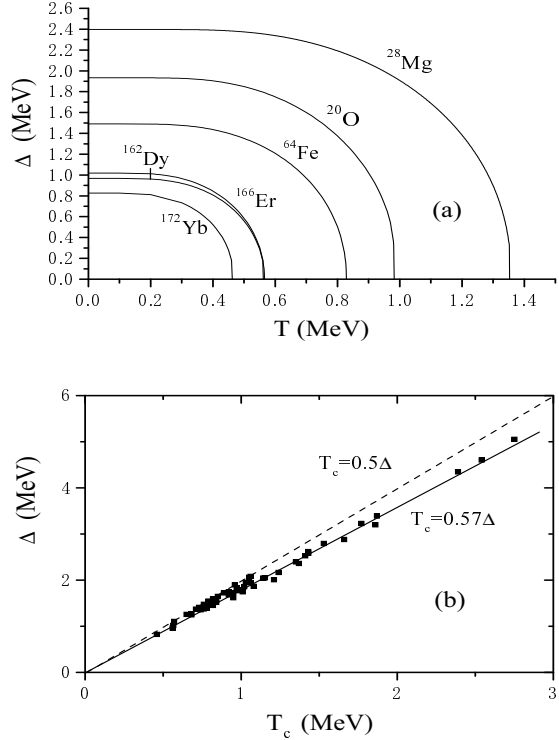}
  \caption{The upper graph (a) shows the neutron pairing gap $\Delta$ versus temperature 
  $T$. The lower graph (b) shows the neutron pairing gap at zero temperature versus the critical 
  temperature $T_{c}$  in the BCS calculation.}
  \label{fig5}
\end{center}
\end{figure}
\begin{figure}[t]
\begin{center}
\includegraphics[width=8cm,height=10cm]{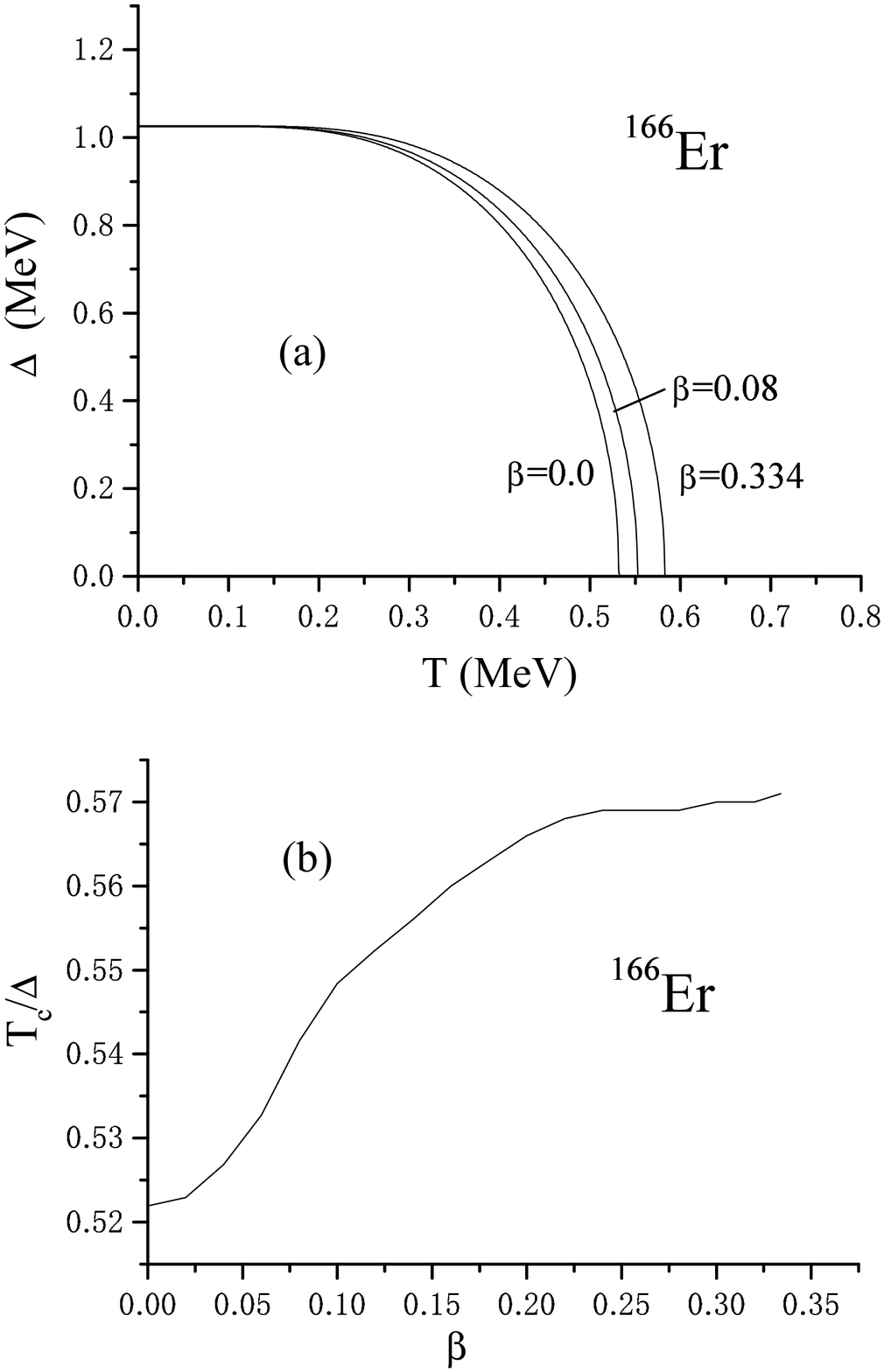}
  \caption{Deformation dependence of the critical temperature $T_{c}$ in the 
  BCS calculation. The upper graph (a) shows the neutron gap in $\beta$=0.0, 0.08, 
  and 0.334. The lower graph (b) shows the ratio of the critical temperature $T_{c}$ 
  and the gap $\Delta_{n}$. }
  \label{fig6}
\end{center}
\end{figure}

We have found correspondence between the "transition 
temperature" and the BCS critical temperature in light nuclei.
It is worth examining whether or not the correspondence obtained in light nuclei
also holds in heavy nuclei over a wide range.
Carrying out exact shell-model calculations is, however, possible only
for light nuclei but not for heavy nuclei. 
We apply the ``correspondence formula" found in the BCS approximation
for the Mg and Ne isotopes to heavy nuclei. 
Since the transition 
temperature $T_{t}$ can be regarded as the critical temperature $T_{c}$ from the 
above discussions, 
it is now interesting to investigate the neutron gap and the critical temperature in the 
BCS calculation solving the gap equation (\ref{eq:6}) with the number equation (\ref{eq:7}) 
over a wide range of even-even nuclei, i.e., $^{20,22}$O, 
$^{20-26}$Ne, $^{24-32}$Mg, $^{28-32}$Si, $^{32-38}$S, 
$^{36-40}$Ar, $^{42-46}$Ca, $^{44-48}$Ti, $^{48-52}$Cr, 
$^{54-66}$Fe, $^{104-124}$Sn, $^{108-114}$Te, $^{124,130}$Ce, 
$^{132}$Nd, $^{134}$Sm, $^{140,144,154}$Gd, $^{162}$Dy, $^{166}$Er, and $^{172}$Yb. 
The pairing force strength $G$ is determined so as to reproduce the experimental 
odd-even mass difference at zero temperature. 
The neutron pairing gap vanishes at the critical temperature for each nuclei as seen 
in the upper graph (a) of Fig. \ref{fig5}. 
The critical temperatures of $^{162}$Dy, $^{166}$Er, and $^{172}$Yb are around 
$T_{c}\approx 0.5$ MeV, which corresponds to 
the peak 
of the experimental heat 
capacity \cite{Schiller,Melby}. 
In addition, the critical temperature $T_{c}\sim 0.8$ MeV for $^{64}$Fe agrees with 
the temperature of peak in the heat capacity obtained using 
the SMMC calculations \cite{Rombouts,Liu}. 
These results support the conjecture that the pairing transition temperature $T_{t}$ corresponds 
to the critical temperature $T_{c}$. 
In the lower graph (b) of Fig. \ref{fig5}, the neutron gaps at zero temperature are plotted 
as a function of $T_{c}$ and the two lines are 
denoted as a guide eye. One is the line $T_{c}\approx 0.57\Delta$ derived from the BCS theory 
\cite{BCS}. The other is that of a simple pairing model with half-filled degenerate shells, 
where the pairing gaps vanish at $T_{c}\approx 0.50\Delta$ MeV \cite{Goodman1}. 
Almost all the plots lie between these two lines, and 
those for deformed nuclei are very close to the line 
$T_{c}= 0.57\Delta$. For instance, the critical temperature of $^{166}$Er exhibits 
$T_{c}=0.58$ MeV where $\Delta$=1.02 MeV and the deformation $\beta$=0.334. 
Since a well deformed nucleus can be regarded as a system with almost uniform level density and 
a small average level spacing compared with the gap $\Delta$ ($\approx$ 1 MeV for the rare 
earth nuclei) \cite{Ring}, this result is reasonable. 
This is simply the case for the metal superconductors. 
The critical temperature is given by $T_{c}\sim 0.57\Delta$ 
$(\Delta=\omega {\rm e}^{-1/\rho G})$ 
where $\omega$ is the phonon energy and $\rho$ the average level density at the Fermi surface, 
and $T_{c}$ is sensitive to the level density. 
On the other hand, 
the simple pairing model with half-filled degenerate shells leads to 
$T_{c}\approx 0.50\Delta$ MeV \cite{Goodman1}. 
Thus it seems that the critical temperature depends on the degree of deformation. 
In fact, Fig. \ref{fig6} shows that the critical temperature $T_{c}$ increases with increasing 
deformation in $^{166}$Er. 

\section{Concluding remarks}

In conclusion, we have studied the pairing transition at finite temperature using the shell 
model calculations. We have demonstrated that the thermal odd-even mass difference is a good 
indicator for the pairing transition at finite temperature as well as the usual one at zero 
temperature. 
We suggest that the pairing correlations can be estimated from the measured level densities 
of nuclei with neutron number $N+1, N,$ and $N-1$, for instance, $^{170}$Yb, $^{171}$Yb, and 
$^{172}$Yb.
It was shown that the transition temperature $T_{t}$ corresponding to 
the inflection point of the curve $\Delta_{n}^{(3)}$ 
is almost identical to the critical temperature $T_{c}$ in the BCS method. 
The pairing correlations almost vanish at two points on the transition temperature $T_{t}$. 
The critical temperature $T_{c}$ depends on the deformation of 
a nucleus, and increases with increasing deformation. The critical temperature $T_{c}$ 
of deformed nuclei follows $T_{c}\approx 0.57\Delta$ MeV as the case of the 
metal superconductors. 
The transition temperature in the case of $N=Z$ nuclei is comparatively higher than 
those of the neighboring $N>Z$ nuclei. 
The {\it p-n} pair correlations seems to contribute to the increase of the 
transition temperature. In addition, as the {\it p-n} pairing is crucial for formation 
of an $\alpha$-like correlated structure in the $N=Z$ nuclei \cite{hasegawa}, 
we can expect the breaking of {\it p-n} pairs as well as like-nucleon pairs 
at a certain temperature. 
Further studies in this direction are in progress. 




\begin{thebibliography} {99}

\bibitem{Bohr} A. Bohr, B. R. Mottelson, and D. Pines, Phys. Rev. {\bf 110} (1958) 936. 
\bibitem{BCS} J. Bardeen, L. N. Cooper, and J. R. Schrieffer, Phys. Rev. {\bf 108} 
             (1957) 1175.
\bibitem{Belyaev} S. T. Belyaev, Mat. Fys. Medd. Dan. Vid. Selsk. {\bf 31}, No. 11(1959).
\bibitem{Sano} M. Sano and S. Yamasaki, Prog. Theor. Phys. {\bf 29} (1963) 397.
\bibitem{Goodman} A. L. Goodman, Nucl. Phys. {\bf A352} (1981) 45.
\bibitem{Ring} P. Ring and Schuck, 'Text book', (1981).
\bibitem{Goodman1} A. L. Goodman, Nucl. Phys. A {352} (1981) 30; Nucl. Phys. A {352} (1981) 45.
\bibitem{Rossignoli} R. Rossignoli, N. Canosa, and P. Ring, Phys. Rev. Lett. {\bf 80} 
                    (1998) 1853.
\bibitem{Schiller} A. Schiller, A. Bierve, M. Guttormsen, M. Hjorth-Jensen, F. Ingebretsen, 
                   E. Melby, S. Messelt, J. Rekstad, S. Siem, and S. W. $\O$degard, Phys. 
                   Rev. C {\bf 63} (2001) 021306R.
\bibitem{Melby} E. Melby, L. Bergholt, M. Guttormsen, M. Hjorth-Jensen, F. Ingebretsen, 
                S. Messelt, J. Rekstad, A. Schiller, S. Siem, and S. W. $\O$degard, 
                Phys. Rev. Lett. {\bf 83} (1999) 3510. 
\bibitem{Rombouts} S. Rombouts, K. Heyde, and N. Jachowicz, Phys. Rev. C {\bf 58} (1998) 3295.
\bibitem{Liu} S. Liu and Y. Alhassid, Phys. Rev. Lett. {\bf 87} (2001) 022501. 
\bibitem{Alhassid} Y. Alhassid, G. F. Bertsch, and L. Fang, Phys. Rev. C {\bf 68} (2003) 044322. 
\bibitem{Cwoik} S. Cwoik, J. Dudek, W. Nazarewicz, J. Skalski, and T. Werner, Comput. Phys. 
                Commun. {\bf 46} (1987) 379. 
\bibitem{USD} B. A. Brown and B. H. Wildenthal, Annu. Rev. Phys. Part. Sci. {\bf 38} (1988) 29. 
\bibitem{Bohr1} A. Bohr and B. R. Mottelson, Nuclear Structure, Vol 1 (Benjamin, NY, 1969). 
\bibitem{Satula} W. Satula, J. Dobaczewski, and W. Nazarewicz, Phys. Rev. Lett. {\bf 81} 
                (1998) 3599.
\bibitem{Dobaczewski} J. Dobaczewski, P. Magierski, W. Nazarewicz, W. Satula, 
                      and Z. Szyma$\acute{\rm n}$ski, 
                      Phys. Rev. C {\bf 63} (2001) 024308. 
\bibitem{Duguet} T. Duguet, P. Bonche, P.-H. Heenen, and J. Meyer, Phys. Rev. C {\bf 65} 
                 (2001) 014311.
\bibitem{kaneko} K. Kaneko and M. Hasegawa, Phys. Rev. C {\bf 60} (1999) 024301.
\bibitem{hasegawa} M. Hasegawa and K. Kaneko, Phys. Rev. C {\bf 61} (2000) 037306.
\end{thebibliography}
\end{document}